\documentclass[proof]{pasj00}

\SetRunningHead{High Performance Parallel Image Reconstruction for New Vacuum Solar Telescope}{Li et al. 2014}

\usepackage{times}

\begin{document}

\title{High Performance Parallel Image Reconstruction for New Vacuum Solar Telescope}
\author{Xue-bao Li\altaffilmark{1,2}, Zhong Liu\altaffilmark{1,2}, Feng Wang\altaffilmark{1,2,3}, Zhen-yu Jin\altaffilmark{1,2}, Yong-yuan Xiang\altaffilmark{1,2}, Yan-fang Zheng\altaffilmark{1,2}}
\altaffiltext{1}{Yunnan Observatories, Chinese Academy of Sciences, Kunming, China, 650011}
\email{lixuebao@ynao.ac.cn}
\altaffiltext{2}{University of Chinese Academy of Sciences, Beijing, China, 100049}
\altaffiltext{3}{Computer Technology Application Key Lab of Yunnan Province, Kunming University of Science and Technology, Chenggong, Kunming, China, 650500}


\maketitle

\begin{abstract}
Many technologies have been developed to help improve spatial resolution of observational images for ground-based solar telescopes, such as adaptive optics (AO) systems and post-processing reconstruction. As any AO system correction is only partial, it is indispensable to use post-processing reconstruction techniques. In the New Vacuum Solar Telescope (NVST), speckle masking method is used to achieve the diffraction limited resolution of the telescope. Although the method is very promising, the computation is quite intensive, and the amount of data is tremendous, requiring several months to reconstruct observational data of one day on a high-end computer. To accelerate image reconstruction, we parallelize the program package on a high performance cluster. We describe parallel implementation details for several reconstruction procedures. The code is written in C language using Message Passing Interface (MPI) and optimized for parallel processing in a multi-processor environment. We show the excellent performance of parallel implementation, and the whole data processing speed is about 71 times faster than before. Finally, we analyze the scalability of the code to find possible bottlenecks, and propose several ways to further improve the parallel performance. We conclude that the presented program is capable of executing in real-time reconstruction applications at NVST.
\end{abstract}

\section{Introduction}
\label{sect:intro}
Atmospheric turbulence is the primary barrier to obtain diffraction limited observation in modern large aperture ground-based solar telescopes, as it randomly distorts the wavefronts radiating from the Sun. The results are image motion, blurring and geometrical distortions in the collected images. Adaptive optics (AO) systems have been introduced to many advanced solar telescopes to enhance spatial resolution of observational images \citep{rimm04,von03,mari10}. However, any AO system only corrects partial wavefront aberrations. To achieve diffraction limited observation, further post-processing reconstruction techniques become indispensable. The most commonly used methods for solar image reconstruction are speckle masking \citep{weig77,lohm83,knox74,von93}, phase diversity methods \citep{gons82,paxm92,lofd94}, and different variants of multiframe blind deconvolution \citep{van98,lofd02,van05}.

The 1-meter New Vacuum Solar Telescope (NVST) has been built at the Fuxian Solar Observatory (FSO) of the Yunnan observatories in 2010. Its main scientific goal is to observe the fine structures in both the photosphere and the chromosphere. Multi-channel high resolution imaging system, with two photosphere channels and one chromosphere channel, has been installed and come into use at NVST. In Table 1, the parameters of acquiring data from the current imaging system of NVST are described. The band for observing the chromosphere is H$\alpha$ (6563 {\AA}), and the band for observing the photosphere is TiO (7058 {\AA}) and G-band (4300 {\AA}), respectively. Because of its channel separation, the imaging system simultaneously acquires data using three detectors, capable of generating 2560$\times$2160 or 1024$\times$1024 pixels image data at a frame rate of around 10 images per second. When observing several hours per day, imaging system will produce several terabytes of unreduced data. Although the storage cost is dropping continually, such huge data volume still becomes difficult to transfer and distribute. The raw image data from NVST are reconstructed using speckle masking. In general, a single ``speckle burst" with at least 100 short exposure images is used to reconstruct one image statistically. The image reconstruction program was originally developed in the Interactive Data Language (IDL). Through an IDL implementation, speckle masking reconstruction of one 2560$\times$2160 pixels image from one burst took about an hour on an Intel Core 3.40 GHz computer, and reconstruction of a single day's data took about several months. Processing such data at least in near real-time can improve the efficiency of the telescope, because not only the data volume is reduced by a factor of 100 rapidly, but also the time between the observations and data analysis is dramatically shortened. Thus, there is a strong need for processing observational data in real-time or at least near real-time. The rapid development of computer technology, particularly in the field of multi-core processors, makes it possible to reconstruct massive speckle data in real-time on site when using speckle masking. Some aspects of using post-processing methods to reconstruct speckle data in near real-time have already been exploited by \cite{denk01}, \cite{woge08} and \cite{woge10}.

In this study, we present a parallelized implementation of speckle masking algorithm. Due to large volume of intermediate data produced and high computational complexity in such a speckle reconstruction, the program is implemented in a multi-processor, multi-node computing environment using the Message Passing Interface (MPI,\cite{grop98}). The remainder of this paper is organized as follows. In Section 2, we make a survey of previous work in the literature concerned with massive speckle data computing. In Section 3, we present system design and a concrete implementation in parallel processing mode. In Section 4, reconstruction results using speckle masking are presented, and the scalability of the code with increasing number of employed processors on a cluster is analyzed followed by discussions and conclusions in Section 5.
\begin{table}
\centering
\caption{The parameters of acquiring data from multi-channel high resolution imaging system of NVST.}
\begin{tabular}{cccc}
\hline
   Channel            &H$\alpha$     &TiO-band    &G-band   \\ \hline
Wavelength ({\AA})   &6563          &7058        &4300      \\
Resolution (pixels)         &1024$\times$1024   &2560$\times$2160     &2560$\times$2160     \\
Frames per second (fps)     &10  &10  &10    \\
Storage capacity (GB per hour)   &75            &398            &398    \\ \hline
\end{tabular}
\end{table}

\section{Related Work}
Among post-processing image reconstruction techniques, the speckle masking becomes attractive not only to NVST but also to other solar telescopes, due to its reliable performance of phase reconstruction. A high resolution image is usually reconstructed from one burst with at least 100 short exposure images. Since speckle masking is only valid for a small region, the isoplanatic patch, the raw images have to be divided into a number of partially overlapping data cubes of 100 subimages \citep{liu14}. Since these subimage cubes are processed independently, this makes a parallel processing of the subimage cubes easy as they are sent to different processors using MPI. The isoplanatic patch has a size of approximately 5''$\times$5''. We use spectral ratio technique to calculate the seeing parameter $r_{0}$ that is necessary for reconstruction \citep{von84}. The modulus of the object's Fourier transform is calculated using classical method of \cite{labe70}. To derive the phases of the object's Fourier transform, we use speckle masking method \citep{weig77}. Inverse transformation of the modulus and phases of the object's Fourier transform yields a mosaic of partially overlapping reconstructed subimages. Finally, all the reconstructed subimages are aligned and put together to form an entire high resolution image. In Figure 1, the schematic diagram of reconstructing a high resolution image is shown, on multiple processors using speckle masking algorithm for NVST.

To deal with both massive speckle data and large amount of computation in near real-time, the technologies of parallel and distributed computing have been applied at many advanced solar telescopes. The Dutch Open Telescope (DOT) reconstructed a full day's data within 24 hours on a cluster consisting of 70 processors \citep{bett04}. Reconstruction of one 1024$\times$1024 pixels image from one burst with 100 images took about 22 seconds, using speckle masking, on a cluster with 23 computation nodes at Dun Solar Telescope (DST) \citep{woge08}. The computation time was 5-6 minutes to reconstruct one 2024$\times$2024 pixels image from one burst with 100 images, using speckle masking on a cluster with 8 computation nodes at New Solar Telescope (NST) \citep{cao10}. The Advanced Technology Solar Telescope (ATST) may meet the goal of reconstructing 80 images of 4096$\times$4096 pixels within 3 seconds into a single image, using speckle masking on a cluster of less than 50 Graphics Processing Units (GPU) \citep{woge12}. We also have developed a faster data handling system to reconstruct images in real-time on a high performance computing cluster for NVST, which is different from that of other telescopes, although parallel implementation of speckle masking reconstruction has been employed at some solar telescopes.

\section{System Design and Implementation}

\subsection{System design}
The use of MPI to implement code for parallel processing has certain implication for the entire design of a program. Because much time has been devoted to reconstructing one 2560$\times$2160 pixels image from one burst, reconstruction of a single burst needs to be accelerated. In general terms, the master process takes charge of not only distributing the tasks to all other processes, receiving the computing results from them, but also performing the calculations as well as them. The overview of the system design is shown in Figure 2, and the pseudo code of the whole parallel implementation of speckle masking is shown in Figure 3. The system shows the whole parallel process of speckle masking on a high performance cluster, assuming a burst of 100 images of 2560$\times$2160 pixels. The image reconstruction application for NVST can be seen as a pipeline mainly consisting of the following procedures: flatfiled and darkfield preprocessing, correlation tracking, estimation of seeing, reconstruction of subimages, and mosaicing of reconstructed subimages. These procedures of image reconstruction are parallelized and accelerated, and the accelerated results are presented in Section 4. In what follows, details of parallel implementation for the individual reconstruction steps are described.

\subsection{Implementation}
In our MPI implementation, we assume the number of processes used in the computation is fixed by the user. However, considering that the number of subimage cubes of one burst in the TiO band is about one thousand, and the number of available processors on the cluster is limited to several hundreds, 222 processes are launched and execute the program on 222 processors in parallel using MPI. The master process reads the average flatfield and darkfield images from the storage device, and broadcasts these images to other processes through two calls MPI$\_$BCAST, assuming the data prepared for preprocessing. The first 100 processes are chosen to read one burst with 100 raw images of 2560$\times$2160 pixels from the storage device based on their process identification number, and perform flatfield and darkfield processing simultaneously and independently. During the correlation tracking, the master process extracts about 2000$\times$2000 pixels subimage from the preprocessed image as a reference image, and broadcasts it to other processes to compensate for image motion. Each of 100 processes computes the image shifts using a two-dimensional cross-correlation method, and executes the same alignment procedure on different images.

We use the spectral ratio technique to calculate the seeing parameter $r_{0}$ from the observed data itself \citep{von84}. A burst of aligned images of 2560$\times$2160 pixels is divided into 80 non-overlapping subimage cubes consisting of 256$\times$256 pixels, different from many partially overlapping subimage cubes as mentioned in Section 2. The scheme of image segmentation for calculating the seeing parameter is shown in Figure 2. The first 80 processes are selected and gather corresponding non-overlapping subimage cubes rapidly through 80 calls MPI$\_$GATHER, via high speed Infiniband network. The data volume of these subimage cubes is small, so the time cost of data transfer between the processes can be neglected. When the data transfer is finished, the first 80 processes execute the same procedure of computing $r_{0}$ on different subimage cubes simultaneously. The number of non-overlapping subimage cubes is less than that of all processes, thus the selected 80 processes can accomplish the tasks of computing the seeing parameters at a time. These seeing parameters calculated by different processes, are collected and averaged by the master process through a call MPI$\_$GATHER. Finally, a mean value of seeing parameter $r_{0}$ is broadcasted to all processes through a call MPI$\_$BCAST.

We also divide a burst of aligned images of 2560$\times$2160 pixels into 1110 partially overlapping subimage cubes consisting of 256$\times$256 pixels, matching the isoplanatic size, in the module of division of all overlapping subimage cubes. The scheme of image segmentation for reconstructing subimage is also shown in Figure 2. All these subimage cubes are assigned and gathered by all processes through many calls MPI$\_$GATHER as indicated in Figure 2 and Figure 3. Each process is designed to reconstruct one subimage from one subimage cube as indicated in Figure 1 and Figure 2, on the premise that it has sufficient amount of RAM memory. However, the number of all processors is insufficient to reconstruct all subimages at a time. The most effective way is to utilize several do-loops to manipulate all subimage cubes on all processes. In the reconstruction of subimages, the object's Fourier modulus is reconstructed according to the classical method of \cite{labe70}, and the object's Fourier phase is reconstructed using speckle masking algorithm \citep{weig77,lohm83}. The same procedure of reconstruction of object's Fourier modulus and phase is implemented by all processes on different partially overlapping subimage cubes simultaneously. Once the operation above is completed, all processes resume simultaneous execution of the same inverse Fourier transform instructions.

The object's Fourier modulus is reconstructed according to the classical method of \cite{labe70}:
\begin{eqnarray}
\mathrm{|O(f)|^{2}}&=&\frac{<|I_{i}(f)|^{2}>_{i}}{<|H_{i}(f)|^{2}>_{i}}.
\end{eqnarray}
$<|H_{i}(f)|^{2}>_{i}$ is generally referred to as the speckle transfer function (STF), which is obtained using spectral ratio technique \citep{von84}. The object's spatial power spectrum $|O(f)|^{2}$ becomes accessible from the measured mean spatial power spectrum $<|I_{i}(f)|^{2}>_{i}$. Once $|O(f)|^{2}$ is calculated, the object's Fourier modulus can be obtained. Due to the lack of possibility to simultaneously observe a reference point source when observing the Sun, the object's Fourier modulus needs to be calibrated with the model STF. In order to choose the correct model function, \cite{woge08a} and \cite{woge08} used the spectral ratio technique \citep{von84} to estimate the strength of atmospheric turbulence, the value of seeing parameter $r_{0}$. In our parallel implementation, we also use the same method to get the value of seeing parameter.

We use several do-loops to reconstruct the subimages on 222 processes, the do-loop index varying from 1 to 5. After 222 subimages are reconstructed in each do-loop, they are aligned simultaneously on different processes. Subsequently, they are collected by the master process, and merged to form a partially full recovered image as illustrated in Figure 2, through a call MPI$\_$REDUCE. A full mosaic high resolution image of 2368$\times$1920 pixels is reconstructed until all do-loops operations are completed. Eventually, the master process writes final results into the storage device, including a full recovered image and corresponding image header information. The program is so intelligent that the important parameters are computed automatically, such as do-loop index, the number of subimage cubes, etc. The do-loop index of reconstructing all subimages can be reduced by using more than 222 processes.

\section{Results and Analysis}
The observations were made with the NVST at FSO without AO on 2 November, 2013. A TiO filter ($\lambda$=705.8nm) was used for observations of solar granulations and sunspots. We used a high speed CCD camera with 16-bit digitization as a detector. Figure 4(a) shows one observed frame of one burst before reconstruction. The field of view is 100$\times$80 $arcsec^{2}$, and the number of pixels is 2368$\times$1920 (0.042 arcsec/pixel). Figure 4(b) shows the speckle masking parallel reconstruction image with the same field of view. The diffraction limited resolution of the telescope is about 0.18 arcsec at $\lambda$=705.8nm. Thus, the speckle reconstructed image is close to the diffraction limited resolution of the NVST. Such reconstructed data can be used to study the dynamical properties of fine features in the photosphere. The image reconstruction program for speckle masking was first developed and carried out in the IDL. Because of the interpreted nature of IDL, the program does not run very efficiently. On a high-end computer equipped with Intel Core(TM) CPU i7-3770 with 3.40 GHz clock speed and 32 GB of RAM, it requires about an hour to reconstruct a 2368$\times$1920 pixels image from a burst of 100 images consisting of 2560$\times$2160 pixels.

The parallel reconstruction has been tested, on a high performance cluster with 45 computation nodes installed at the Yunnan Observatories, while not changing the results. The high performance cluster we used consists of 45 computer servers. Each server has 2-way Intel Xeon E5-2650 with 2.00 GHz CPUs (total 16 cores) and 64 GB of DDR3 RAM. All nodes are connected together with a high-speed Infiniband network with bandwidth of 40 Gb/s. The Red Hat 6.2 (64-bits) operation system is installed on each server. OpenMPI-1.6.4 architecture is installed on each node. The storage system, a Lustre distributed file system with four servers, is also connected with 40 Gb/s Infiniband network. The total storage capacity is 54 TB. All raw data are firstly copied to the storage system of the cluster for high performance data processing. For the tests, we used a burst of 100 images, and this data set was reconstructed with 184775040 bispectrum values and 1110 subimage cubes consisting of 256$\times$256 pixels.

Table 2 shows the runtime comparisons of reconstructing a 2368$\times$1920 pixels image between various modules in C language using MPI on 222 processors of the cluster, and IDL on the high-end computer as mentioned above. Since the IDL implementation cannot run on one node of the cluster, we used the high-end computer with almost the same computing capacity in place of it. Several time-consuming module operations as listed in Table 2 are successfully parallelized, and the speedup shows positive results. The runtime of the whole data parallel processing reduces to around 48 seconds when using 222 processors, and the whole data processing speed is around 71 times faster than before. The most time-consuming module of reconstruction of all subimages shows a great speed increase by a factor of about 202, the module of flatfield and darkfield preprocessing is about 193 times faster than before, and the processing speed of other modules is about 9-41 times faster than before. In the module of division of all overlapping subimage cubes, about 16 seconds are used to transfer data between all processes. Although the data volume of 1110 subimage cubes is around 55 GB, the total amount of data transfer increases to 92 GB. The total amount of data transfer increases when the number of processors increases as shown in Figure 5(c). This reason is that division of all overlapping subimage cubes generates additional transfer data using more than 100 processes, when one burst with 100 aligned images is stored in the first 100 processes. Because the number of do-loops for reconstructing all subimages is 5 on 222 processors, 92 GB of total transfer data is also divided into 5 groups. Thus each group in each do-loop has around 18.4 GB of transfer data, avoiding the jam of network. The procedure of data transfer is shown in detail in the pseudo code of Figure 3. The communication time makes up a significant proportion of one image reconstruction time. If we reduce the amount of data transfer and increase the bandwidth of network, the performance of parallel reconstruction will be further improved, which is what many efforts we should focus on in our future work.

We have assessed the scalability of the parallel implementation by measuring the time used for computation of a speckle reconstruction. The execution time for the parallel implementation between various modules on the cluster is shown in Figure 5(a) and Figure 5(b) for the number of employed processors ranging from 32 to 555. In the curves of all modules in Figure 5(b), the runtime reduces when the number of processors increases. However, the speedup is not linear especially when the number of processors used exceeds 111, and there is a saturation at a minimum. Because only the first 100 processes can perform parallel processing in all modules in Figure 5(b), the runtimes wouldn't change obviously when the number of processors varies from 111 to 555. In the curve of all subimages reconstruction in Figure 5(a), the runtime always reduces when the number of processors increases, because the number of do-loops for reconstructing all subimages reduces. However, the runtime always increases in the curve of division of all overlapping subimage cubes in Figure 5(a) when the number of processors increases, because the total amount of data transfer increases. In the curve of the whole data processing in Figure 5(a), the runtime reduces when the number of processors increases, and there is a saturation at a minimum of around 48 seconds when the number of processors exceeds 222. This reason is that the runtime increase in the module of division of all overlapping subimage cubes almost equals the runtime decrease in the module of all subimages reconstruction. In any event, already today a system such as the one tested above has achieved real-time performance, for a detector that reads out and stores a 2560$\times$2160 pixels frame at an effective rate of 10 frames per second.

\begin{table}
\centering
\caption{The runtime comparisons of reconstructing a 2368$\times$1920 pixels image between various modules in IDL and C language using MPI.}
\begin{tabular}{cccc}
\hline
   Module            &C $\&$ MPI (seconds)     &IDL (seconds)    &Factor   \\ \hline
Flatfield and darkfield preprocessing    &0.13          &25.2        &193.8      \\
Correlation tracking          &7.5   &139.6     &18.6     \\
Estimation of seeing     &3.5  &31.4  &9    \\
Division of all overlapping subimage cubes   &16.4  &-   &-      \\
Reconstruction of all subimages &15.3  &3100 &202.6   \\
Mosaicing of all reconstructed subimages  &2.5   &104.1   &41.6  \\
The whole data processing   &48            &3425            &71.3    \\ \hline
\end{tabular}
\end{table}

\section{Discussion and Conclusions}
In this study, we present the details of parallel implementation for image reconstruction using speckle masking in solar application. The code is written in C language for parallel processing using MPI, and the parallel implementation between various modules shows a great speed increase as compared to previous IDL implementation. The spatial resolution of reconstructed image can be improved significantly compared to that of original image. We also tested the scalability of the code. Timing results of parallel implementation between various modules on the cluster showed a clear advantage with greater numbers of processors. Since considerable proportion of the computing time is spent in the modules of division of all overlapping subimage cubes and reconstruction of all subimages, we need to not only increase the network bandwidth, but also accelerate subimage reconstruction, to further improve performance of the parallel reconstruction.

Obviously, high performance image reconstruction like speckle masking will be valuable for NVST and next generation solar telescopes when a high performance cluster is adopted on site, as it increases the telescope's efficiency. In addition, the development of multiprocessing and parallel numerical algorithms for high spatial resolution imaging will become even more important in the context of advanced large aperture solar telescopes, when both the chip size and data acquisition speed of the detector increase. The program proposed in the study has been successfully tested on the cluster, and the outstanding real-time performance has been achieved. However, we still try to seek new techniques for real-time image reconstruction for NVST.

General Purpose Graphics Processing Units (GPGPU) is a new parallel computing technique that has been widely used in real-time computing. After the implementation of MPI image reconstruction, we also consider to migrate our program from MPI to GPU in order to obtain higher computing speed. In the preliminary experiment, we used one GPU to accelerate bispectrum calculation, and gain significant speedup (around 4-6 times than that of MPI) performance enhancement in one subimage reconstruction. However, the massive data transfer from host memory to GPU memory is a time-consuming issue because of the limitation of computer's bus bandwidth. The optimal reconstruction algorithm on GPU is worth to study in the future.

\section*{Acknowledgements}
This work is supported by the National Natural Science Foundation of China (No. U1231205, 11163004, 11203077). The authors thank the NVST team for the support they have given this project, and also gratefully acknowledge the helpful comments and suggestions of the reviewers.

\begin{figure}
\begin{center}
\includegraphics[width=1.0\textwidth]{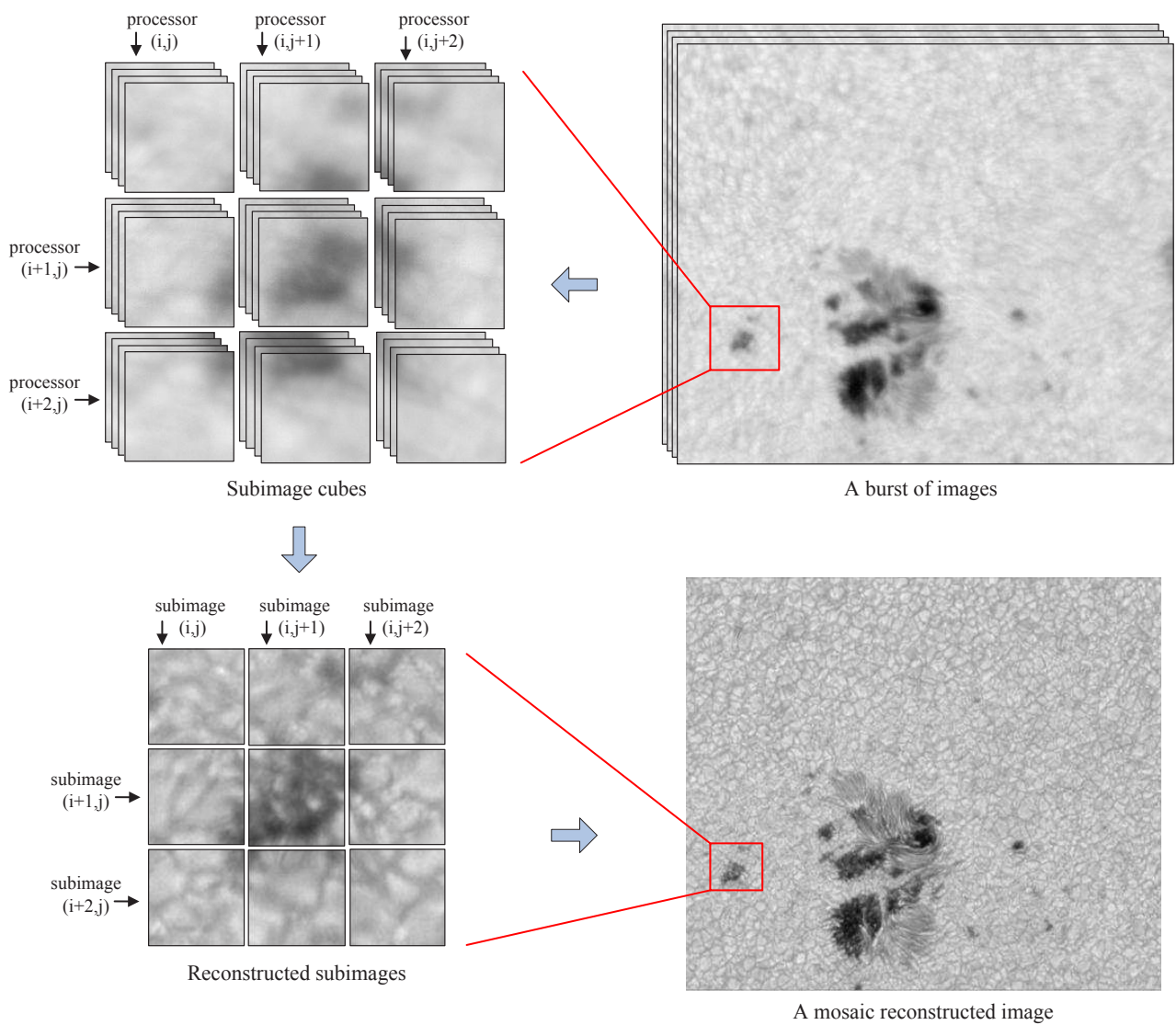}
\end{center}
\caption{The schematic diagram of reconstructing a high resolution image, on multiple processors using speckle masking algorithm.}
\label{figure1}
\end{figure}

\begin{figure}
\begin{center}
\includegraphics[width=0.43\textwidth]{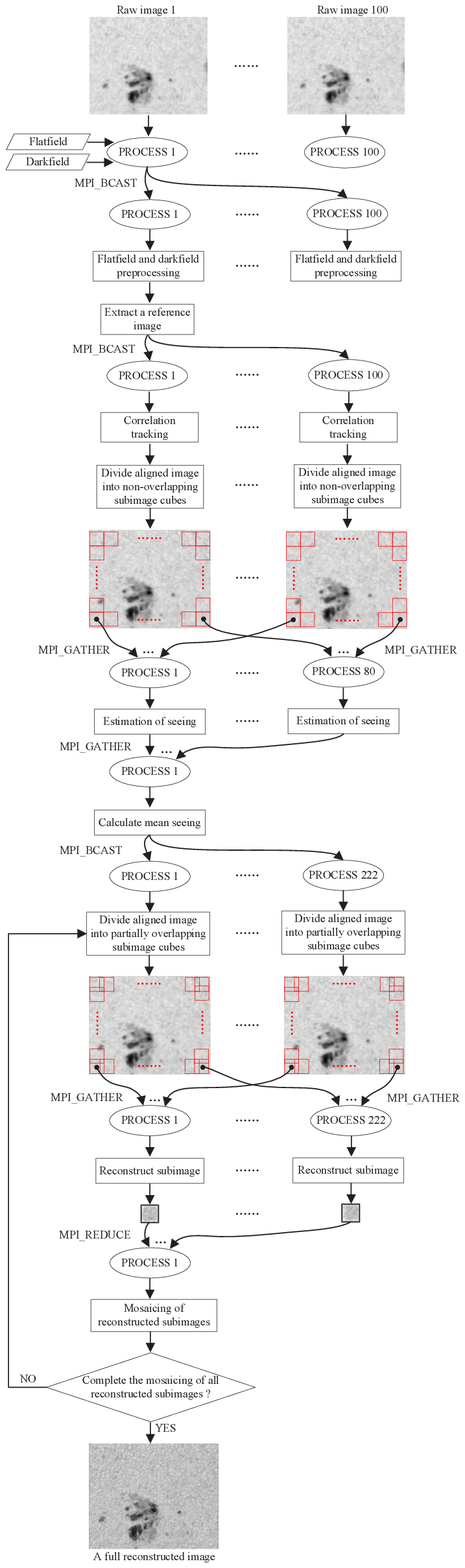}
\end{center}
\caption{ The overview of the system design. The system shows the whole parallel process of speckle masking on a high performance cluster, assuming a burst of 100 images of 2560$\times$2160 pixels. }
\label{figure2}

\end{figure}

\begin{figure}
\begin{center}
\includegraphics[width=0.9\textwidth]{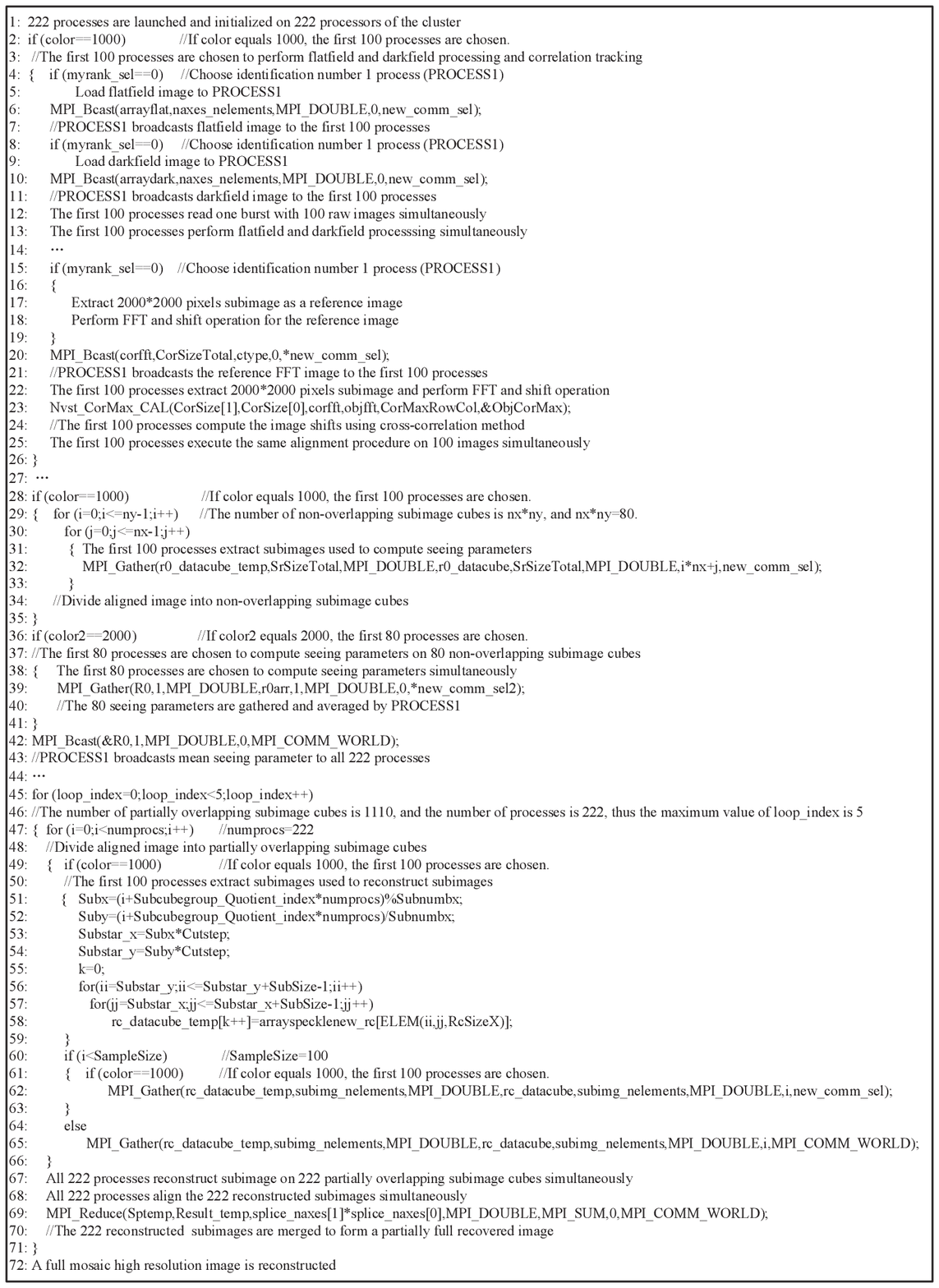}
\end{center}
\caption{ Pseudo code of the whole parallel implementation of speckle masking on a high performance cluster, assuming a burst of 100 images of 2560$\times$2160 pixels. }
\label{figure2}

\end{figure}

\begin{figure}
\begin{center}
\includegraphics[width=0.9\textwidth]{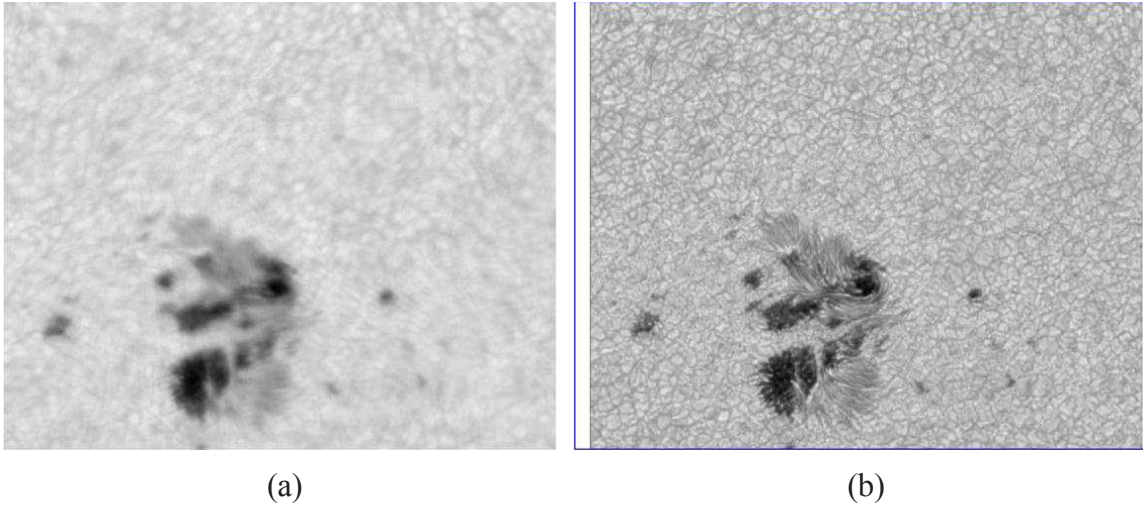}
\end{center}
\caption{ (a) One observed frame of one burst before reconstruction. (b) The speckle masking parallel reconstructed image. The field of the view is 100$\times$80 $arcsec^{2}$. }
\label{figure2}

\end{figure}

\begin{figure}

\begin{center}
\includegraphics[width=0.9\textwidth]{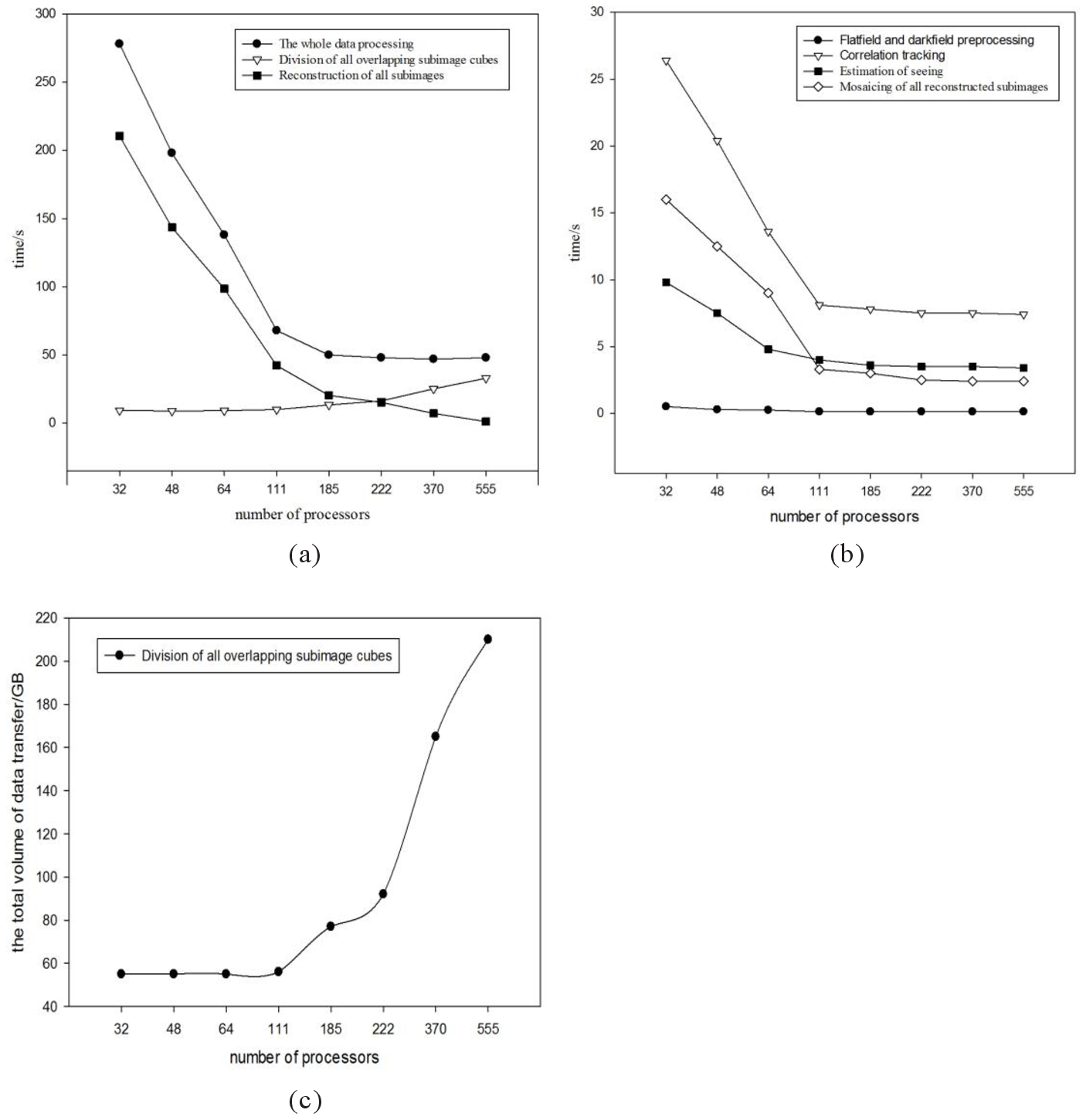}
\end{center}
\caption{ Time used for one reconstruction between various modules ((a) and (b)) and volume of data transfer (c) versus number of employed processors on the high performance cluster. }

\end{figure}


\begin{thebibliography}{00}

\bibitem[Bettonvil et al.(2004)]{bett04}
Bettonvil, Felix C. M., et al. 2004, SPIE, 362, 373

\bibitem[Cao et al.(2010)]{cao10}
Cao, W. D., et al. 2010, SPIE, 77355V1

\bibitem[Denker et al.(2001)]{denk01}
Denker, C., Yang, G., \& Wang, H. 2001, Solar Physics, 63, 70

\bibitem[Gonsalves(1982)]{gons82}
Gonsalves, R. A. 1982, Optical Engineering, 829, 832

\bibitem[Gropp \& Snir(1998)]{grop98}
Gropp, W., \& Snir, M. 1998, MPI the Complete Reference, MIT Press, Cambridge

\bibitem[Knox \& Thompson(1974)]{knox74}
Knox, K. T., \& Thompson, B. J. 1974, ApJ, L45, L48

\bibitem[Labeyrie(1970)]{labe70}
Labeyrie, A. 1970, A\&A, 85, 87

\bibitem[Liu et al.(2014)]{liu14}
Liu, Z., Xu, J., Gu, B. Z., et al. 2014, RAA, 705, 718

\bibitem[Lofdahl(2002)]{lofd02}
Lofdahl, M. G. 2002, SPIE, 146, 155

\bibitem[Lofdahl \& Scharmer(1994)]{lofd94}
Lofdahl, M. G., \& Scharmer, G. B. 1994, SPIE, 254, 267

\bibitem[Lohmann et al.(1983)]{lohm83}
Lohmann, A. W., Weigelt, G., \& Wirnitzer, B. 1983, Applied Optics, 4028, 4037

\bibitem[Marino et al.(2010)]{mari10}
Marino, J., W\"{o}ger, F., \& Rimmele, T. 2010, SPIE, 77363E-1

\bibitem[Paxman et al.(1992)]{paxm92}
Paxman, R. G., Schulz, T. J., \& Fienup, J. R. 1992, JOSA, 1072, 1085

\bibitem[Rimmele et al.(2004)]{rimm04}
Rimmele, T. R., Richards, K., Hegwer, S., Fletcher, S., Gregory, S., \& Moretto, G. 2004, SPIE, 179, 186

\bibitem[van Kampen \& Paxman(1998)]{van98}
van Kampen, W. C., \& Paxman, R. G. 1998, SPIE, 296, 307

\bibitem[van Noort et al.(2005)]{van05}
van Noort, M., Rouppe van der Voort, L., \& Lofdahl, M. G. 2005, Solar Physics, 191, 215

\bibitem[von der L\"{u}he(1984)]{von84}
von der L\"{u}he, O. 1984, JOSA, 510, 519

\bibitem[von der L\"{u}he(1993)]{von93}
von der L\"{u}he, O. 1993, A\&A, 374, 390

\bibitem[von der L\"{u}he et al.(2003)]{von03}
von der L\"{u}he, O., Soltau, D., Berkefeld, T., \& Schelenz, T. 2003, SPIE, 187, 193

\bibitem[Weigelt(1977)]{weig77}
Weigelt, G. P. 1977, Optics Communications, 55, 59

\bibitem[W\"{o}ger \& Ferayorni(2012)]{woge12}
W\"{o}ger, F., \& Ferayorni, A. 2012, SPIE, 84511C-1

\bibitem[W\"{o}ger et al.(2010)]{woge10}
W\"{o}ger, F., Uitenbroek, H., Tritschler, A., et al. 2010, SPIE, 773521-1

\bibitem[W\"{o}ger \& von der L\"{u}he(2008)]{woge08a}
W\"{o}ger, F. \& von der L\"{u}he, O. 2008, SPIE, 70191E-1

\bibitem[W\"{o}ger et al.(2008)]{woge08}
W\"{o}ger, F., von der L\"{u}he, O., \& Reardon, K. 2008, A\&A, 375, 381

\end{thebibliography}
\end{document}